\documentclass[12pt]{article}
\usepackage{epsf}
\usepackage{cite}
\def\beq{\begin{eqnarray}}
\def\eeq{\end{eqnarray}}

\def\Mpl{M_{\rm Pl}}
\def\Rbr{R}
\def\Rbu{{\cal R}}
\def\M{M_*}
\def\gn{G_N}
\def\tgn{{\tilde G}_N}

\begin{document}
\begin{flushright}
NYU-TH/01/05/15 \\
TPI-MIN-01/23 \\
UMN-TH-2009/01\\
hep-th/0106001
\end{flushright}

\vspace{0.1in}
\begin{center}
\bigskip\bigskip
{\large \bf  Nonperturbative 
Continuity in Graviton Mass\\
\vspace{0.1in}
versus Perturbative Discontinuity}

\vspace{0.4in}      

{C\'edric Deffayet$^a$, Gia Dvali$^a$, Gregory Gabadadze$^b$, 
and Arkady Vainshtein$^b$}
\vspace{0.1in}

{\baselineskip=14pt \it 
$^a$Department of Physics, New York University,  
New York, NY 10003\\[1mm]
$^b$Theoretical Physics Institute, 
University of Minnesota, Minneapolis, MN 55455\\} 
\vspace{0.2in}
\end{center}

\vspace{0.9cm}
\begin{center}
{\bf Abstract}
\end{center}
We address the question whether a graviton could have a small 
nonzero mass.  The issue is subtle for two reasons:
there is a discontinuity in the mass in the lowest tree-level
approximation, and, moreover, the nonlinear four-dimensional 
theory of a massive graviton is not defined unambiguously.
First, we reiterate the old argument that 
for the vanishing graviton mass the lowest tree-level approximation 
breaks down since the higher order corrections are singular in 
the graviton mass. However, there can exist nonperturbative solutions
which correspond to the summation of the singular terms,
and these solutions are continuous in the graviton mass.  
Furthermore, we study a completely nonlinear 
and generally covariant five-dimensional model 
which mimics the properties of the four-dimensional 
theory of massive gravity.
We show that the exact solutions of the model are 
continuous in the mass, yet the perturbative 
expansion exhibits the discontinuity in the leading order
and the singularities in higher orders 
as in the four-dimensional case. Based on exact cosmological solutions
of the model we argue that the helicity-zero graviton state which 
is responsible for the  perturbative discontinuity  
decouples from the matter in the limit of vanishing 
graviton mass in the full classical theory.

\newpage

\section{Introduction}

Could a  graviton be massive? 
The naive answer to this question seems to be  
positive. Indeed, if the graviton Compton 
wavelength, $\lambda_g\,=\,m_g^{-1}$, is large enough, 
let us say of the 
present Hubble size, we should  not be able to tell 
the massive graviton from a  massless one.
In fact, astrophysical bounds  are even milder, $\lambda_g > 10^{24}
{\rm cm}$  \cite {PDG} (see also Refs. \cite {Will}). 
However, in general relativity (GR) the issue turns out to 
be more subtle. A dramatic observation has been made in 
Refs. \cite {Iwa,Veltman,Zakharov} according to which
predictions of massless GR, such as the
light bending by the Sun 
and the precession of the Mercury perihelion, differ
by numerical factors from the predictions of the theory with a
massive graviton,  no matter how small the  graviton mass is. 
This discontinuity, if true, would unambiguously prove that graviton is 
strictly massless in  Nature.

The arguments of Refs. \cite {Iwa,Veltman,Zakharov} were based on the 
lowest tree-level approximation to 
interactions between sources. In this approximation 
the discontinuity has a clear physical interpretation. 
Indeed, a massive graviton in four-dimensions  has {\it five} 
physical  degrees of freedom (helicities $\pm 2$, $\pm 1$, $0$) while the
massless graviton has only two (helicities $\pm 2$). The exchange by  the three
extra degrees of freedom  can be interpreted in the limit $m_g\to 0$ as an
additional contribution due to  one 
massless vector particle with two degrees of
freedom  (``graviphoton'' with helicities $\pm 1$) plus one real scalar
(``graviscalar''  with the helicity 0).
The graviphotons do not contribute to the one-particle 
exchange --- their derivative coupling to
the conserved energy-momentum tensor vanishes. 
The graviscalar, on the other hand,  is coupled to the trace of the 
energy-momentum tensor and its  
contribution is generically nonzero. It is what causes the 
discontinuity between the predictions of massless 
and massive theory in the lowest tree-level approximation. 

However, as was argued in Ref.~\cite {Arkady},
this discontinuity does not persists
in the full classical theory.
It was shown that 
the lowest tree-level approximation 
to the calculation of interactions between two sources 
breaks down when graviton mass is small.  
The next-to-leading terms in the corresponding expansion 
are huge since they are 
inversely proportional to powers of $m_g$.  
Thus, the truncation of the perturbative series  
does not make much sense and all higher order terms in the 
solution of classical
equations for the graviton field should be summed up.
The summation leads to the nonperturbative solution which is continuous
when $m_g\to 0$. The perturbative discontinuity shows up only at large
distances where higher order terms are small, these distances 
are growing when $m_g\to 0$. In other words, the continuity 
is not perturbative and not uniform as a function of distance.

A simple reason why one could expect 
the violation of the lowest tree-level approximation 
is that it does not take 
into account the characteristic physical scale of the problem; 
while the nonperturbative calculation of the Schwarzschild 
solution does account for this effect.
In the nonperturbative solution 
the coupling of the extra scalar mode to the matter 
is suppressed by the ratio of graviton mass to the 
physical scale of the problem. 
Hence, the predictions of the massive theory could be made
infinitely close to the predictions of the massless theory
by taking small $m_g$.

The argumentation can be conveniently 
presented by considering  the gravitational amplitude of scattering of a probe
particle in the background gravitational field produced by a heavy static
source. This amplitude has the following generic structure  (note, that we use
the flat metric $\eta_{\mu\nu}={\rm diag}(-1,1,1,1)$): 
\beq
{\tilde h}^{\mu\nu}(q)~{t}^{\,\prime}_{\mu\nu}~\propto~
{ a(q^2)\;{t}^{\mu\nu}\,{t}^{\,\prime}_{\mu\nu}
-b(q^2)\;{t}^{\mu}_{\mu}\;{t}^{\,\prime\,\nu}_\nu
\over q^2+m_g^2-i\epsilon}~,
\label{1}
\eeq
where $t_{\mu\nu}=p_\mu p_\nu$ and
$t^{\,\prime}_{\mu\nu}=p^{\,\prime}_\mu p^{\,\prime}_\nu$ refer to the heavy
particle with the four-momentum 
$p_\mu=(M,{\vec 0})$ and to
the light particle with the momentum $p^{\,\prime}_\mu$ 
correspondingly\,\footnote{To avoid the confusion 
note that  we use ${t}_{\mu\nu}$
only as a kinematical structure of the vertices not implying
that it is the energy-momentum tensor.}, see
Fig.~1.
The form factors $a(q^2)$ and
$b(q^2)$ are functions of the momentum transfer $q^2$ and are defined by two
parameters: the graviton mass
$m_g$ and the Schwarzschild radius $r_M=2G_N M$ of the 
heavy particle  with the mass $M$.
\begin{figure}[ht]
\epsfxsize=12cm
\centerline{%
   \epsfbox{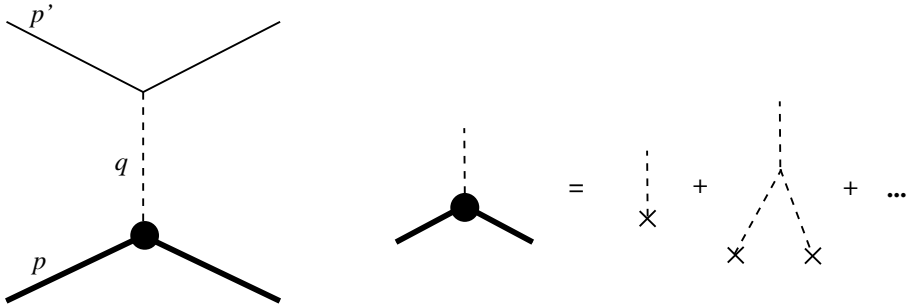}%
         }
\caption{Scattering of the probe particle at the gravitational 
field of the heavy
source. The bold circle accounts for summation 
of the higher order iterations 
over the nonlinearities in the classical equations.} 
\end{figure}

In the lowest tree-level approximation of the massive
theory the form factors $a$ and $b$ are just constants and the 
unitarity (sum over five helicities) fixes their
ratio, $a=3b$,  while the same unitarity with two graviton states (helicities
$\pm 2$)  in the massless theory gives $a=2b$.  Therefore, the discontinuity 
\cite{Iwa,Veltman,Zakharov} appears. 
However, this  is  only valid for the small
momenta $q\ll m_g\,(m_g\,r_M)^{-1/5}$, for which 
the higher order corrections are
small \cite{Arkady}. In the coordinate space it means that the linear
approximation becomes valid only  at the distance 
\begin{equation}
r\gg r_m\,,\qquad r_m\equiv \frac{(m_g\,r_M)^{1/5}}{m_g}=
\frac{(2G_N M m_g)^{1/5}}{m_g}~,
\end{equation}
which for the Sun is bigger that the solar system 
size (see discussions in the next section).

On the other hand, at $q\gg m_g\,(m_g\,r_M)^{-1/5}$, i.e., 
at  shorter distances,
$r\ll r_m$, we expect that 
the summation of higher orders \cite{Arkady} returns the relation
$a=2b$ of the massless theory. In other words, nonperturbative summation
should lead to the decoupling of the graviscalar from the 
heavy source for distances  $r\ll r_m$.

What was not verified in Ref.~\cite{Arkady} is a matching of the 
nonperturbative
solution at $r\ll r_m$ with the exponentially decreasing linear solution
at $r\gg r_m$. 
It might happen indeed that the solution matches
an exponentially increasing function instead\,\footnote{
Such solution can still be acceptable as long as 
the  exponential growth of the solution takes 
over at distances much larger than the observable size of the Universe. 
This will take place if the graviton Compton wavelength  
$\lambda_g\gg 10^{28}~{\rm cm}$.}.
Boulware and Deser in \cite{BD} expressed their
doubts about an existence of the large distance matching. Moreover, they
argued that there is no consistent interacting theory of the massive spin-2 
field
in 3+1 dimensions.  One of the arguments in Ref. \cite{BD} 
was  that at quantum level the theory contained the sixth polarization  
in addition to the standard five polarizations. Furthermore,
the mass term in the action is not uniquely defined
beyond the quadratic order in the fields. 

These legitimate concerns can be addressed by embedding the
4D theory of a massless graviton into a five-dimensional theory --- 
a route we take in the present paper. Indeed, gravity in five
dimension is well defined as a classical 
gauge theory, a massless graviton has exactly
five states. For the matter fields which are confined
to the four-dimensional brane 
the theory mimics the massive spin 2
particle with the fifth component of the momentum playing the role of the
mass\,\footnote{Note the analogy with the supersymmetric BPS states whose
mass is given by a central charge. This charge also can be viewed as an extra
component of the momentum in the
dimensionally enlarged space.}.

The model which we  discuss is that of Ref.~\cite{dgp}.  In this model
matter is
localized on a brane. The brane world-volume theory contains the induced 4D
Einstein-Hilbert term due to which  a five-dimensional graviton
 mimics the massive four-dimensional spin-2 state on the brane. In
contrast with the 4-dimensional massive theory,  in this case the full
nonlinear
action can be written.   The two-body problem for sources on the brane 
is now well-defined. The amplitude has the same generic form (\ref{1})
with substitution of $q^2+m_g^2$ by $q^2+m_c q$, where $m_c$ is a counterpart
of $m_g$ in the model. We present the arguments in favor of aforementioned
behavior of the form factors $a(q^2)$ and $b(q^2)$. However, we did not manage
to obtain the exact solution of the Schwarzschild problem in this case either. 

Instead,
we derive a number of evidences supporting the conjectured behavior from 
the exact cosmological solutions~\cite {Cedric,ddgl} of the model.
We show that the lowest tree-level perturbative result is off by a
factor $4/3$ as compared with the exact result and 
explain why the corresponding perturbation theory breaks down. 
Based  on this,  we expect that the perturbative discontinuity  
is absent  on the
nonperturbative level in the full classical theory indeed. 

Recently the problem of the vanishing 
graviton mass was studied in a different setup.
It was shown in Refs. \cite {Kogan} and \cite
{Porrati} that there is no mass discontinuity even in the 
lowest tree-level exchange on  de Sitter (dS) \cite {Higushi,Kogan}
or Anti de Sitter (AdS) \cite {Kogan,Porrati} backgrounds\,\footnote{
The consideration for the dS space is a bit subtle since
for $m^2_g<2\Lambda /3$ ($\Lambda$ being the cosmological constant)
unitarity is violated in the theory \cite {Higushi}.}.
This fits well with the discussions presented above. Indeed, 
in the case of the (A)dS background, even the lowest tree-level 
approximation does take into account the presence of a 
mass scale of the problem, which in that  case is given by the 
cosmological constant $\Lambda$. It was shown 
in \cite {Porrati} that the coupling
of graviscalar is proportional to $m^2_g/\Lambda$ when $m_g\to 0$,
and deviations from the massless model vanish in this limit.
Since the cosmological constant in our world is restricted as 
$\Lambda \leq 10^{-84}~{\rm GeV}^2$, then
the allowed graviton mass  is in the range $m_g\ll 10^{-42}~{\rm GeV}$ ---
i.e., the  graviton 
Compton wavelength  is bigger than our horizon size. 
The existence of such a tiny 
graviton mass 
is  immaterial for all astrophysical and cosmological 
observations \cite {Kogan}
(see also an interesting discussion of the continuity issue in the
recent work~\cite {Kogan1} ). Note, that the nonperturbative continuity
allows for much wider range for the graviton mass, $m_g\ll (r_M/r^5)^{1/4}$.
Here $r$ is the maximal distance from the Sun where the data
are obtained, see Sec.~2 for numerics.

In Ref.~\cite {Duff} it was argued that in the (A)dS background the 
perturbative discontinuity reappears at the one-loop quantum level ---
the  phenomenon very similar to the one-loop discontinuity
for a massive non-Abelian vector fields discussed in \cite{Veltman}.
This is certainly true since the loops are sensitive to 
the number of particles running in the loop diagrams. From the practical 
point of
view, however, the comparison of the theory with the  experimental data on
the  light bending by the Sun  and the precession of the Mercury perihelion is
not affected by the small quantum loop  corrections. Indeed, while the 
graviscalar decouples from the classical source it is still coupled to
the graviton
and  does contribute to the 
quantum loops.  However, such effects of quantum gravity are suppressed and 
most likely  cannot be disentangled in solar system measurements.  For these
reasons,  in what follows we are 
focusing on the (dis)continuity in the classical theory only.

The paper is organized as follows. In Sec.~2 we recall 
the essence of the  graviton mass 
discontinuity found in Refs. \cite {Iwa,Veltman,Zakharov} and
discuss the results of Ref.~\cite {Arkady} 
where it was shown that there is in fact the 
continuity in the graviton 
mass in the full classical theory.
In Sec.~3 we introduce the five-dimensional nonlinear 
model which mimics the properties of a four-dimensional 
massive gravitational theory. We show that the perturbative 
discontinuity which is present
in the lowest tree-level approximation disappears in the 
exact solution of the model. In Sec.~4 we discuss
another exact solution of the nonlinear model which
interpolates between the four-dimensional 
and five-dimensional regimes. We conclude in Sec.~5.   
 
\section{Preliminaries: Massive Graviton in 4D}

We will consider the following action for a massive graviton 
on a flat 4D background:
\beq
S_m\,=\, \Mpl^2\int d^4x\,\sqrt{|g|}\, \left (    
R\,+\, {m_g^2\over 4}\left[h_{\mu\nu}^2\,-\,(h^\mu_\mu )^2\right]
\right ) \,,
\label{PF}
\eeq
where $g_{\mu\nu}=\eta_{\mu\nu}+h_{\mu\nu}$
and the Planck mass $\Mpl$  is related to the Newton constant $\gn$ as
$\Mpl^2=1/(16\pi\,\gn)$. The
mass term has the Pauli-Fierz form 
 \cite {Pauli}, in quadratic in $h_{\mu\nu}$ terms it is the only form
which does not  introduce  ghosts \cite {Neu}. We imply that indices 
 in the mass term are raised and lowered by
the tensor $\eta_{\mu\nu}$. If it were $g_{\mu\nu}$ instead the difference
would appear only in the cubic and higher in  $h_{\mu\nu}$ 
terms which are not fixed anyway;  
higher powers of $h_{\mu\nu}$  could be  arbitrarily 
added to the mass term. 

In order to see the presence of the discontinuity 
in the lowest tree-level approximation let us compare free graviton 
propagators in the massless and massive theory.
For the massless graviton we find:
\beq
D^0_{\mu\nu ;\alpha\beta}(q)\,=\, \left(
{1\over 2} \,\eta_{\mu\alpha}  \eta_{\nu\beta}+
{1\over 2} \, \eta_{\mu\beta}  \eta_{\nu\alpha}-   
{1\over 2} \, \eta_{\mu\nu}  \eta_{\alpha\beta}\right)\frac{1}{
q^2-
i\epsilon}\,,         
\label{4D}
\eeq  
where only the momentum independent parts
of the tensor structure is kept. By a gauge  choice the momentum dependent
structures can be taken to be zero.
On the other hand, there is no gauge freedom for the massive gravity given 
by the action (\ref{PF}), and the
propagator   takes the following form:
\beq
D^m_{\mu\nu ;\alpha\beta}(q)\,=\, 
\left(
{1\over 2} \,\tilde\eta_{\mu\alpha} \tilde \eta_{\nu\beta}+
{1\over 2} \, \tilde\eta_{\mu\beta}  \tilde\eta_{\nu\alpha}-   
{1\over 3} \, \tilde\eta_{\mu\nu} \tilde \eta_{\alpha\beta}\right)\frac{1}{
q^2+m_g^2-
i\epsilon}\,,
\label{5D}
\eeq  
where 
\begin{equation}
\tilde\eta_{\mu\nu}=\eta_{\mu\nu}+\frac{q_\mu q_\nu}{m_g^2}\,.
\end{equation}
Note the $1/m_g^4$, $1/m_g^2$ singularities of the propagator. 

The difference in the  numerical coefficients for the $\eta_{\mu\nu} 
\eta_{\alpha\beta}$ structure in the massless and massive propagators (1/2
versus 1/3) is what leads to the perturbative 
discontinuity \cite {Iwa,Veltman,Zakharov}.
No matter how small the graviton mass
is, the predictions are substantially  different 
in the two cases. The structure (\ref {5D}) gives 
rise to contradictions with observations.

To see how this comes about let us 
calculate the amplitude
of the lowest tree-level exchange by a graviton  between two 
sources with energy-momentum tensors $T_{\mu\nu}$
and  $T^{\prime}_{\alpha\beta}$ (the sign tilde denotes the quantities
which are Fourier transformed to momentum space):
\beq
{\cal A}_0\,\equiv\,-8\pi\,\gn\,{\tilde T}_{\mu\nu}\,D_0^{\mu\nu;\alpha\beta}
\,{\tilde T}^{\prime}_
{\alpha\beta}\,=\,- \frac{8\pi\,\gn}{q^2}\left 
( {\tilde T}_{\mu\nu}\,-\,{1\over 2}\,
\eta_{\mu\nu} 
\,{\tilde T}^\beta_\beta   \right )\, {\tilde T}^{\prime\mu\nu}\,.
\label{4DT}
\eeq
In the massive case this amplitude takes the form:
\beq
{\cal A}_m\,\equiv\,-8\pi\,\gn\,{\tilde T}_{\mu\nu}\,D_m^{\mu\nu;
\alpha\beta}
{\tilde T}^{\prime}_
{\alpha\beta}\,=\, -\frac{8\pi\,\gn}{q^2+m_g^2}\left ( {\tilde
T}_{\mu\nu}\,-\,{1\over 3}\,
\eta_{\mu\nu} 
\, {\tilde T}^\beta_\beta   \right )\, {\tilde T}^{\prime\mu\nu}\,.
\label{4DTm}
\eeq
In the relativistic normalization we are using 
${\tilde T}_{\mu\nu}= 
\langle p|{T}_{\mu\nu}|p \rangle=2p_\mu
p_\nu$ at zero momentum transfer, $q=0$.
Suppose we take two probe massive static sources with masses $M_1$ and
$M_2$. Then only $\tilde T_{00}$, $\tilde T^\prime_{00}$ are non-vanishing
and the lowest tree-level graviton exchange   determines the Newtonian
interaction,
\begin{eqnarray}
&&V_0(r)=\int\! {{\rm d}^3 q\over (2\pi)^3} \,{\rm e}^{i{\vec q} 
{\vec r}}\,\frac{{\cal
A}_0}{4M_1 M_2}\,=\,- \frac{\gn M_1 M_2}{r}\,, \nonumber\\[1mm]
&&V_m(r)=\int\! {{\rm d}^3 q\over (2\pi)^3} \,{\rm e}^{i{\vec q} {\vec
    r}}\, \frac{{\cal A}_m}{4M_1 M_2}\,=\, 
-\frac{4}{3}\;\frac{\gn M_1 M_2}{r}\,{\rm e}^{-m_g r}~.
\label{poten}
\end{eqnarray}
Expressions (\ref {4DT}) and (\ref {4DTm})
give different results for the Newtonian attraction even in the range $r\ll
\lambda_g$ where one can neglect the exponential decrease. This
difference can be eliminated by redefining  the Newton  coupling for the
massive  theory as follows:
\beq
\tilde\gn\,=\,{4\over 3}\,\gn\,\,,
\label{tgn}
\eeq
where $\gn$ is the Newton  constant  of the massless theory. For 
nonrelativistic problems the  predictions of the massive theory 
with  
the coupling rescaled by a factor $3/4$ at $m_g\to 0$ are identical to 
those of the massless theory with the coupling $\gn$.

However, this is not enough to warrant
the viability of the massive model. The relativistic predictions 
in the two cases are different \cite {Veltman,Zakharov}. 
For instance, the predictions for  the light bending by the Sun
are in conflict.
At the classical level the trace of the 
energy-momentum tensor for light is zero. Therefore,
the second term  on the right hand side of Eqs.~(\ref {4DT})
and (\ref {4DTm}) is not operative for light. 
Hence,  the amplitudes 
${\cal A}_0$ and ${\cal A}_m$
are identical in this case. However, we have established above 
that calculations in the massive theory should be 
performed with the rescaled Newton constant. 
Taking into account this fact,
the prediction for the light bending in the massive theory
is off by 25$\%$ \cite {Iwa,Veltman,Zakharov}. 
 
We could certainly take an opposite point of view.
Namely, do not rescale the Newton constant of the massive theory.
In this  case the predictions for the light bending in the 
massive and massless models are identical. However, the 
Newton force between static sources would  
differ by a factor of $4/3$.

The above considerations are based on the lowest
perturbative  approximation. The question is whether 
these results hold in the full classical theory. Normally, 
one would expect that for the solar system distances the lowest  
approximation is well justified. 
However, it was argued in Ref. \cite {Arkady}
that the approximation breaks down in the  massive theory
for relatively short distances. 
Since this breaking manifests itself in a rather interesting way 
we will briefly summarize  the results of Ref. \cite {Arkady} below.

To see the inconsistency of the perturbative expansion in $\gn$
let us look (following \cite {Arkady}) at the Schwarzschild
solution of (\ref {PF}). We parametrize the 
interval for a massive spherically symmetric  body as follows:  
\beq
ds^2\,=\,-{\rm e}^{\nu (\rho)}\,dt^2\,+{\rm e}^{\sigma(\rho)}\,d\rho^2+
{\rm e}^{\mu(\rho)}\,\rho^2\,(d\theta^2+{\rm sin}^2\theta\,d\phi^2)\,.
\label{interval}
\eeq
In the massless  theory the function $\mu$ is redundant
due to the reparametrization invariance of the theory;
it can be put equal to zero. 
However, in the massive  case this gauge symmetry is broken 
and $\mu$ is nonzero. Therefore, 
in order to compare the results in the massive and  
massless case  one has to do the substitution: 
\beq
r\,\equiv\, \rho \,{\rm exp} \left ({\mu\over 2}  \right )\,, 
\qquad
{\rm exp} \left ( \lambda  \right )\,\equiv \, 
\left (1+{\rho\over 2} {d\mu \over d\rho}  \right )^{-2} 
{\exp} \left (\sigma -\mu \right )\,.
\label{lambda}
\eeq

The standard Schwarzschild solution 
of the massless theory takes the following form:
\vspace{0.5cm}
\begin{eqnarray}
&&\nu^{\rm Schw}(r) \,=\, -
\lambda^{ \rm Schw}(r)\,=\, \ln\left(1- {r_M\over
r}\right)\,=\, - {r_M\over r}- \frac 1 2 \left( {r_M\over
r}\right)^2+\ldots\;,\nonumber\\[1mm]
&& \mu^{\rm Schw} (r)\,=\, 0\;.
\label{Schw}
\end{eqnarray}
Here $r_M\equiv 2\gn M$ is the gravitational radius of the 
source of mass $M$. 

Let us compare this with the perturbative in $\gn$ solution
of the massive theory obtained in Ref.
\cite {Arkady}. In the leading plus  next-to-leading approximation
in $\gn$ the solution reads:
\begin{eqnarray}
&& \nu\,\simeq \, -{r_M\over r}\left [1+{7\over 32}\,
{r_M\over m_g^4\,r^5}\right ]\,,
\nonumber\\[1mm]
&&
\lambda\,\simeq \,{1\over 2}\, 
{r_M\over r}\left [1-{21\over 8}\,{r_M\over m_g^4\,r^5}\right ]\,,
\nonumber\\[1mm]
&&
\mu\,\simeq \,{1\over 2}\, 
{r_M\over m_g^2\, r^3}\,\left [1\,+\,{21\over 4}\,{r_M\over m_g^4\,
r^5}\right ]\,.
\label{mum}
\end{eqnarray}

We note the following peculiarities of results (\ref{mum}):
\begin{itemize}

\item{In the leading order 
there is the finite discontinuity in the expression for 
$\lambda$: the result of the massless theory in (\ref{Schw})
differs from the result of the massive model by a factor $1/2$.
This is precisely the discontinuity which is seen in the 
lowest approximation.}

\item{The next-to-leading corrections in
(\ref{mum}) are governed by the ratio $ {r_M/ m_g^4r^5}$
and are singular in $m_g$.}

\item{For any given distance $r$ there is a 
value of $m_g$ below which  the perturbative expansion in 
$\gn$  breaks down.} 

\end{itemize}

These results are in correspondence with the perturbative series 
for the scattering
amplitude described by Feynman graphs in Fig.~1. The leading terms in the 
expansions (\ref{mum}) are given by the diagram of the
first order in the source, i.e., the diagram with one cross.  The singular in
$m_g$ terms in the propagator (\ref{5D}) do not contribute in this order.
In the next order (the diagram with two crosses in Fig.~1) we have two
extra propagators which could provide a singularity in $m_g$ up to $1/m_g^8$.
Two leading terms $1/m_g^8$ and $1/m_g^6$ do not contribute 
again so the result contains 
only the $1/m_g^4$ singularity  as in Eq. (\ref{mum}).

To demonstrate how badly the expansion in powers
of $\gn$ breaks down let us take the  largest allowed 
value for the graviton mass, 
$m_g=(10^{25}~{\rm cm})^{-1}$ \cite {PDG,Will}  
and calculate the correction to the leading result 
in the gravitational field of the Sun.
We will find that at distances of order of  the solar system size, i.e.,
at $r\sim 10^{15}~ {\rm cm}$, the next-to-leading 
corrections in (\ref{mum}) are about $10^{32}$ 
times bigger than the leading terms. 
Therefore, this expansion is unacceptable.

For a light enough graviton, however, 
a consistent perturbative expansion could be
organized in powers of $m_g$. In this case one finds \cite {Arkady}:  
\begin{eqnarray}
&&\nu(r) \,=  \,-{r_M\over r}+{\cal O}\left(m_g^2\sqrt{r_M
r^3}\right)\,,\qquad
\lambda(r) \,= \,{r_M\over r}+{\cal O}\left(m_g^2\sqrt{r_M
r^3}\right)\,,\nonumber\\[1mm]
&& \mu(r)\, = \,\sqrt{8\,r_M\over 13\, r}+{\cal O}\left(m_g^2\,
r^2\right)\,,
\label{mg}
\end{eqnarray}
where only the leading terms in $r_M/r$ are retained. These expressions 
are valid  in the following interval:
\beq
r_M\,\ll\, r \,\ll\, r_m\,,\qquad r_M=2\gn M\,,\qquad r_m={(m_g\,
r_M)^{1/5}\over m_g }\,.
\label{int}
\eeq
For the gravitational field of the Sun this 
would correspond to the interval:
\beq
3\cdot 10^{5}\,{\rm cm}\,\ll \, r \,\ll \,10^{21}\,{\rm cm}\,,
\label{intn}
\eeq
where the lower bound is less than the radius of the Sun
and the upper bound is of the order of a  galaxy scale. 
Thus, for practical calculations within the  
solar system this expansion is well suited.

As we see,  the expressions for 
$\nu$ and $\lambda$ in the leading approximation 
coincide with those of the massless theory
(\ref {Schw}). 
Thus, there is {\it no mass discontinuity}.
Moreover,  the expressions (\ref{mg}) explicitly 
shows non-analyticity in $\gn$, 
$\mu\propto \sqrt{\gn}$, for $\nu$ and $\lambda$ non-analytic terms are
proportional to $m_g^2$. 

We discussed in the Introduction subtle  issues  
concerning the validity of the results discussed above
arising even on the classical level: the nonlinear theory of massive 
gravity is not uniquely 
defined and it is complicated to make sure 
that the solutions which have no discontinuity    
do indeed satisfy the boundary conditions
at infinity, i.e., that  for  $r\gg1/m_g$ the solution  matches
the exponentially decreasing function.

As we already noted even the exponentially growing solution 
can be acceptable when the graviton  Compton wavelength
becomes larger than the observable size of the Universe. 
The Yukawa factors due to the graviton mass, $\exp (\pm m_g r)$,
can be made to be arbitrarily close to the unity by decreasing the 
graviton mass. However, as we discussed above, this does
not warrant the continuity of the $m_g \to 0$ limit
since  the coefficients in front of the perturbative 
potentials in the massive and massless theory (\ref{poten})
are different and $m_g$ independent. 
Therefore, the question whether the 
graviton could  have a nonzero mass, effectively reduces to the 
question whether the graviton could have
five polarizations. Indeed, these extra polarizations are
responsible for the $m_g$ independent 
discontinuity in the coefficients in the potentials (\ref{poten}). 
Therefore, in what follows below we will address the question:
{\it ``Can the graviton which describes the data in our observable 
Universe have five degrees of freedom?''}

In the next section  we  present a model based on five dimensions
where massless graviton naturally has five degrees of freedom.
The model is free of all the problems of the 4D massive gravity
discussed above. We perform our  analysis
within this completely nonlinear theory  in which exact solutions
can be found. These solutions are  compared with  the 
perturbative results. We find that the picture outlined in the 
work \cite {Arkady} (and discussed above) holds.

\section{A Brane Model of Massive Graviton}

The 5D model which we will discuss was introduced in 
\cite {dgp}. The gravitational part of the 
action takes the form:
\beq
S~=~M_*^3 ~ \int d^4x ~dy~ \sqrt{|G|}~ \Rbu 
 +~
\Mpl^2 ~ \int d^4x ~ \sqrt{|g|}~\Rbr(x)~. 
\label{action}
\eeq
Where $M_*$ is a parameter of the theory and $\Mpl=1.7\cdot 
10^{18}~{\rm GeV} \gg M_*$.
Furthermore, $G_{AB}$ is 5D metric tensor, $A\!=\!\{\mu, 5\}\!=
\!\{0,1,2,3,5\}$,  and $\Rbu$ is the five-dimensional Ricci scalar, 
$g_{\mu\nu}$ denotes the induced metric on the brane
which we take as 
\beq
g_{\mu\nu} (x)~\equiv~G_{\mu\nu}(x, y=0)~,\qquad \mu,\nu=0,1,2,3~, 
\label{gind1}
\eeq
neglecting the brane fluctuations.

We assume that our observable 4D world (4D matter) 
is confined to a tensionless brane (a tensionless 
hyper-plane in this case) which is fixed at the point 
$y=0$ in extra fifth dimension\,\footnote{A simplest possibility is 
to consider a brane at a fixed point of the ${\bf R}/{\bf Z}_2$
orbifold.}.  In other words, 
we assume that the energy-momentum tensor 
of 4D matter has the following factorized form 
$T_{\mu\nu}(x)\,\delta(y)$. 
We also imply the presence of the Gibbons-Hawking boundary term
on the brane, this provides the correct Einstein equations in the bulk.
These simplifications help to keep the presentation clear and 
do not affect  our main results. The brane world aspects of 
the model (\ref {action}) were studied in detail in 
Refs. \cite {dgp,dg,dgkn,ddg}.

Let us study the gravitational potential between two static 
bodies located on the brane. This   
can be calculated from the action (\ref {action}). The corresponding Green
function is conveniently represented by working in momentum space in the four
world-volume directions and in position space with respect to
the transverse coordinate  $y$. For the time being we can neglect
the tensorial structure of the propagator (to be discussed below)
and calculate the scalar part of the Green function.
This can  be done by calculating the corresponding propagator
in a theory with scalars only which have the 
bulk and brane kinetic terms similar to  
(\ref {action}). The result of the calculation  
reads as follows \cite{dgp}:
\beq
{\tilde G}(q,y=0)\,=\,{1\over \Mpl^2}~
{1\over q^2\,+\,m_c\,\sqrt{q^2}}\,,
\label{prop}
\eeq
where we introduce the parameter 
\beq
m_c\,\equiv\,{1\over r_c}\,\equiv \,{2\,\M^3  \over \Mpl ^2}\,.
\label{co}
\eeq
The Green function (\ref {prop}) 
has unusual features. It has a tachyonic pole 
at $q^2=-q_0^2+{\vec q}^2=m_c^2$ which corresponds to the 
decay into the continuous tower of Kaluza-Klein states (which arise from 
the reduction of 5D graviton).  Although, the five-dimensional graviton
is well defined, from the 4D perspective it looks
as unstable particle with the width $m_c$. Nevertheless,
the rules of integration for the propagator (\ref {prop}) 
in the complex energy plane can be defined consistently. 

In particular, using  (\ref {prop}) we can
find the static potential $\phi(r)$. The result  
can be  written in terms of 
special functions and has different asymptotic behavior for small  and 
large distances (see Ref. \cite {dgp}).
The ``crossover scale'' between these two
regimes is defined by $r_c$ given in Eq.  (\ref {co}).
At short distances, i.e.,  when $r\ll r_c $
\beq
\phi(r)\,=\,-{1\over 8\pi^2 \Mpl ^2}\,{1 \over r}\,\left \{
{\pi\over 2} +\left [-1+\gamma -{\rm ln}\left ( {r_c\over r } \right ) 
\right ]\left ( {r\over r_c } \right )\,+\,{\cal O}(r^2)  
\right \}\,.
\label{short}
\eeq
Here $\gamma\simeq 0.577$ is the Euler constant.
The leading term in this expression has the familiar $1/r$ scaling of the
four-dimensional Newton  law with a right  numerical coefficient.
The leading correction is given by the 
logarithmic {\it repulsion} term in (\ref {short}).   

Let us turn now to the large distance behavior. For $r\gg r_c $ one finds:
\beq
\phi(r)\,=\,-{1\over 16\pi^2 \M^3}\,{1 \over r^2}\,+
\,{\cal O} \left ( {1\over r^3} \right )\,.
\label{long}
\eeq
The long distance potential scales as $1/r^2$ in accordance 
with the  5D Newton law.
Thus, the crossover scale (\ref {co}) should be sufficiently large to
avoid conflict with astronomical observations. In \cite{dgp} it was estimated
that for $M_* \sim$ 1 TeV, the crossover scale $r_c$ 
is around $10^{15}$ cm, which is
roughly the size of the solar system. This is too low to be consistent with
data. Therefore, the scale $\M$ should be taken to be 
at least a couple of orders smaller then 1~TeV. 
This is in no conflict with any gravitational or 
Standard Model measurements (see discussions in Ref. 
\cite {dgkn,ddg}). We take $r_c\ge 10^{25}~{\rm cm}$
which corresponds to $\M\le 1\, {\rm GeV}$.

The parameter $m_c$
plays a role   in this model which  
in many respects is similar to that of the graviton
mass $m_g$ in (\ref {PF}). Indeed,
as $m_c\rightarrow 0$, gravity on a brane becomes 
4D Newtonian at more and more larger  distances.
Moreover, the four-dimensional interaction in the model with 
the action (\ref {action})
can be interpreted as an  exchange of 
a four-dimensional state with the width 
equal to $m_c$ \cite{dgp}. 
In the next section we will find even more 
closer similarities between $m_c$ and $m_g$.

\subsection{Perturbative Discontinuity}

To see that the model (\ref {action}) exhibits the 
discontinuity in the one-graviton tree-level 
approximation let us calculate, 
following \cite {dgp},
the tensorial  structure of one-graviton exchange.
To this end  we will solve the Einstein equations in the 
linear approximation in $h_{AB}$ which is the deviation from the flat 
5D metric,
\beq
G_{AB} ~=~ \eta_{AB} ~+~h_{AB}~.
\label{pert}
\eeq
We choose the {\it harmonic~ gauge} in the bulk: 
\beq
\partial^A h_{AB}~ =~{1\over 2}~ \partial_B h^C_C~.
\label{gauge}
\eeq 
In this gauge from  the $\{\mu 5\}$ and  $\{55\}$ components of the sourceless 
equations of motion follows that
\beq
h_{\mu 5} ~=~0,\qquad ~h^5_5~=~h^\mu_\mu~.
\label{mu5}
\eeq

Let us turn to the $\{\mu\nu\}$ components
of the Einstein equations. After some simplifications 
they take the form:
\beq
\left ( \M^3 \partial_A\partial^A+\Mpl^2 \,\delta(y)
\,\partial_\alpha \partial^\alpha 
\right )h_{\mu\nu} = -\left \{ T_{\mu\nu} -{1\over 3}\eta_{\mu\nu} 
T^\alpha_\alpha   \right \}\delta(y)
+\Mpl^2\,\delta(y) \,\partial_\mu
\partial_\nu \,h^\alpha_\alpha\,.  
\label{basic}
\eeq
There are two  terms on the right hand side of this equation.
The first one has a structure which is identical to that of  
a massive 4D graviton (or, equivalently of a massless 5D graviton).  
The second term on the right hand side which contains derivatives 
$\partial_\mu\partial_\nu$ is not important at the moment 
since it vanishes when it is contracted with the 
conserved energy-momentum tensor.  
As a result, the amplitude of interaction
of two test sources takes the form:
\beq
{\tilde h}_{\mu\nu}(q, y=0)~{\tilde T}^{\prime\mu\nu}(q)~\propto~
{  {\tilde T}^{\mu\nu}{\tilde T}^{\prime}_{\mu\nu}~-~
{1\over 3} ~{\tilde T}^{\mu}_\mu
{\tilde T}^{\prime\nu}_\nu \over q^2~+~m_c\, q }~,
\label{prop1}
\eeq  
where $q\equiv \sqrt{q^2}$. 
We see that the tensor structure 
is the same as in the case of the massive 4D theory, 
see Eq. (\ref {4DTm}).

In analogy with the discussions in the previous section we
could expect that the lowest tree-level approximation 
will break down in the next iterations in 
the classical source. 
Further indication on this is the existence 
of the singular in $m_c$ terms in the expression
for the gravitational field $h_{\mu\nu}$ 
produced by a static source. 
We write the  energy-momentum tensor for the 
source  as follows:
\beq
T_{\mu\nu}(x) \,=\,- M\,\delta_{\mu 0}\,\delta_{\nu 0}\,\delta^{(3)}
(\vec {x})\,,
\label{source}
\eeq
where $M$ is its rest mass. As before, let us 
make Fourier transform with respect to four world-volume 
coordinates.
Then the solution looks as follows:
\beq
{\tilde h}_{00}(q,y)&=& c\,{1\over 2}\,\tgn \,M\,{1\over q^2\,+\,m_c\,q}\,
{\rm exp}\left( -q|y| \right )\,,
\label{00} \\[1mm]
{\tilde h}_{ij}(q,y)&=& c\,{1\over 4}\,\tgn \,M\,{\delta_{ij}
\over q^2\,+\,m_c\,q}\,
{\rm exp}\left( -q|y| \right )\, \nonumber \\[1mm]
&+&c\,\tgn\,M\,{q_i\,q_j\over m_c\,q}\,
{1\over q^2\,+\,m_c\,q}\,
{\rm exp}\left( -q|y| \right )\,, 
\label{ij}
\eeq
where $c=-16\pi$.
These expressions, taken at $y=0$,  should be contrasted
with the lowest order expressions for the Schwarzschild solution
in 4D theory with a massless graviton:
\beq
{\tilde h}^{\rm Schw}_{00}(q)\,=\,c\,{1\over 2}\,G_N\,M\,
{1\over q^2}\,,
\label{00sch}
\eeq
\beq
{\tilde h}^{\rm Schw}_{ij}(q)\,=\,c\,{1\over 2}\,G_N\,M\,{\delta_{ij}
\over q^2}\,.
\label{ijsch}
\eeq

Comparing the expressions 
(\ref {00}-\ref {ij}) to those in (\ref {00sch}-\ref {ijsch}) 
we draw   the following conclusions:

\begin{itemize}

\item{Upon the substitution $\tgn \to  G_N$ the $\{00\}$ components 
coincide for large momenta, or, equivalently  for
$r\ll r_c$.}

\item{The $\{ij\}$ component of the 5D theory
consists of two terms. The first term,
after the substitution $\tgn \to  G_N$  
is twice as small as the corresponding term 
on the right hand side of the Schwarzschild solution (\ref {ijsch}). 
This is what gives rise to the discontinuity.} 

\item{There is an additional term in the 
expression for ${\tilde h}_{ij}(q,y=0)$ which is 
proportional to:
$$
{q_i\,q_j\over m_c\,q}\,.
$$
This term 
does not contribute to the  one-graviton exchange 
in the leading order because of conservation 
of the energy-momentum tensors 
(the diagram with a single cross in Fig.~1).
However, it does contribute to higher order diagrams
(the ones with two and more  crosses in Fig.~1).
This term is singular in $m_c$  
and the perturbation theory
in $\gn$ breaks  down when $m_c\to 0$.}

\end{itemize}

Given these arguments, we conclude that for a consistent calculation
of the interaction between two  sources on a brane 
we should find the Schwarzschild solution 
which sums up all the orders of the Born expansion for 
the classical equations.
Unfortunately, we could not manage to find the analytic solution.
However, implying the existence of a smooth in $m_c\to 0$ limit,
one could perform the expansion in $m_c$ 
in analogy with  the 4D massive case \cite {Arkady}.

The $\{\mu\nu\}$ component of the Einstein equation
for the action (\ref {action}) can be integrated with respect to $y$ 
in the interval $-\epsilon\leq y \leq \epsilon$ 
with $\epsilon \rightarrow 0$. The resulting equation takes the form:
\beq
{\overline {\cal G}}_{\mu\nu}(x)+
m_c\int_{-\epsilon}^{+\epsilon}{\cal G}_{\mu\nu}(x,y)\,dy
\,=\,-{M \over 2~\Mpl^2}~ 
\delta_{\mu 0} \delta_{\nu 0}\delta^{(3)}(x)~,
\label{eqint}
\eeq
where   ${\overline {\cal G}}_{\mu\nu}$ 
and ${\cal G}_{\mu\nu}(x,y)$ 
denote the Einstein tensor
of the worldvolume and bulk theories respectively.
Since the extrinsic curvature has a finite jump  across the 
brane, the  second term on the left hand side of
(\ref {eqint}) is nonzero even in the limit $\epsilon \rightarrow 0$.
This term is proportional to the parameter $m_c$ with respect to which 
the expansion is performed (we imply that the metric is 
nonsingular in $m_c$, this seems to be a reasonable
requirement  for a physically meaningful solution).

Then, it is clear from (\ref {eqint}) that
in the lowest approximation in $m_c$ one recovers the 
usual 4D Schwarzschild solution of the massless theory 
(\ref {Schw}). For the calculation of the sub-dominant 
corrections in $m_c$ and for matching conditions
at infinity, however, numerical simulations are needed.
Note that in this case the solution should  be matched at infinity
to a well known 5D Schwarzschild 
solution which decreases as $(r_M/r)^2$ at infinity. 
This is an easier task compared to the 4D massive case 
where the power-low solution 
at short distances should be matched with  the Yukawa potential 
at infinity\,\footnote{The next morning after this paper
was submitted to the archive an interesting work \cite {Dick}
appeared. In Ref. \cite {Dick} the asymptotic form
of the Schwarzschild solution for $m_c\to 0$ was also 
discussed and, moreover, certain generalizations of 
cosmological solutions of the model (\ref {action}) 
were obtained.}.

Does this mean that we cannot compare analytically
the perturbative  and nonperturbative results  
in the model (\ref {action})? Not at all. 
Instead of finding the exact Schwarzschild solution
we perform the 
similar analysis for other solutions which can be obtained
explicitly. In the 
next section we discuss an exact  nonperturbative 
cosmological solution of  the model (\ref {action}) 
found in Refs. \cite {Cedric,ddgl} 
which differs from the perturbative  result by $4/3$.

\subsection{Nonperturbative continuity}

In this section we study the cosmological solution in the 
model ({\ref {action}) found in Ref.
\cite {Cedric} and \cite {ddgl}. It was already noticed in \cite {Cedric}
that the cosmological evolution in (\ref {action}) 
is governed by the Newton constant which 
differs from the the ``Newton'' constant 
of perturbation theory by $4/3$.
We will discuss in details 
this discrepancy. 

Our goal is  as follows. We consider
the solution of the model (\ref {action}) 
which describes the expansion of the matter dominated Universe. 
We will perform two distinct calculation for 
this. First we find the solution based on
the Newtonian approximation. This calculation makes use of
the lowest order  potential between objects on the brane.
As a second step we find the corresponding exact 
nonperturbative cosmological 
solution of the Einstein equations. In the domain 
where the Newtonian  approximation is legitimate, 
the perturbative result for the cosmological solution  would 
coincide under the normal circumstances 
with the nonperturbative one  as it happens in 
4D world with a massless graviton.
However, we find the discrepancy by a factor
of $4/3$ in these two methods.

Let us start with the perturbative approach.
As we established in the previous subsection
the one-graviton exchange in the lowest approximation 
gives rise to the following expression for 
the potential of a massive source at short distances $r\ll r_c$:
\beq
\phi(r)\,=\,-\tgn\,{M\over r}\,.
\label{nrV}
\eeq
The appearance of the constant $\tgn$ 
instead of $\gn$ in this expression is
related to the fact that we used the lowest tree-level approximation.

Let us now use the standard consideration of the 
Newtonian cosmology\,\footnote{For a careful treatment and interpretation 
of the Newtonian cosmology see, e.g., \cite {Dc}.}.
Consider a spherical ball with some uniform matter density in it.
We assume that the radius of the ball $R$ is much smaller then 
$r_c$ and that we are in a regime where the Newtonian approximation is 
valid. In this case the potential of the ball on its surface 
takes the form:
\beq
\phi_{\rm ball}(r=R)\,=\,-\tgn\,{M\over R}\,.
\label{U}
\eeq
Let us consider a point-like probe particle of mass $m_0$ which is located 
just right on the surface of the ball. We neglect the back-reaction of
this probe particle on the ball. The energy conservation condition
for the system of the ball and probe takes the form: 
\beq
{m_0\,{\dot R}^2\over 2}\,-\,\tgn\,{M\,m_0\over R}\,=\,{k\,m_0\over2}\,,
\label{energy}
\eeq
where dots denote the time derivative and $k$ is some constant. 
We would like to calculate the time evolution of the 
radius $R$.  In the regime which we discuss this is  equivalent to the
time evolution of the scale factor in Friedmann-Lemaitre-
Robertson-Walker cosmology. 
In what follows we consider the solution which 
corresponds to the expansion of flat, i.e., $k=0$, matter-dominated Universe. 
For $k=0$ we rewrite (\ref {energy}) as follows:
\beq
\left ( {  {\dot R}\over R} \right)^2\,=\,{8\pi\over 3}\,\tgn\,
\rho\,,
\label{Friedmann}
\eeq
where the density $\rho$ for the matter-dominated Universe 
is related to the scale factor $R$ as follows
\beq
\rho\,=\,{u\over R^3}\,,
\label{kr}
\eeq
where $u$ is some constant.

This is nothing but the Friedmann equation for the scale factor $R$
for a flat matter-dominated Universe.  We find the solution for the 
scale factor:
\beq
R^3(t)\,=\,6\,\pi \,u \,\tgn\, t^2\,.
\label{R3}
\eeq
This solution is consistent with the fact that we choose the time period when
$R\ll r_c$ so that the brane world evolves in accordance with laws
of 4D theory. What is important in our solution is 
the numerical coefficient in the relation (\ref {R3}) which different from
 the 4D massless gravity case -- it contains $\tgn=(4/3)\gn$ instead of $\gn$.
Below we will show that  the exact solution matches the massless gravity in 
the limit  $m_c \to 0$. 

Before discussing the exact solution let us explain why the Newtonian 
approach outlined above does not produce a correct coefficient. It is
due to effects of nonlinear terms: similar to the Schwarzschild problem
in 4D massive gravity discussed in Section 2 these corrections are
defined by powers of the parameter
\begin{equation}
\frac{G\,u}{m_c^2 \, R^3}\sim \frac{1}{m_c^2 \, t^2}\,.
\end{equation}
It is clear that these corrections blow up at $m_c \to 0$ and we need
to sum them up. The corrections seem to be small  at the later time $t
\gg 1/m_c$, but as we will see the 4D approach stops to work at this
epoch.

Let us now solve the same problem 
using the exact Einstein  equations. 
We parametrize the 5D interval in the following form:
\beq
ds^2\,= \,-N^2(t,y)\,dt^2\, +\, A^2(t,y)\, dx_idx^i\, + \,B^2(t,y) \,
dy^2\,.
\label{inter}
\eeq
The 4D scale factor is defined as follows:
\beq
R(t)\,\equiv\, A(t, y=0)\,.
\label{R}
\eeq
The solution 
was found in \cite {Cedric} and \cite {ddgl}:
\beq
N\,=\,1\,-\,|y|\,{{\ddot R}\over {\dot R}}\,, \qquad 
A\,=\,R\,-\,|y| \,{\dot R}\,, \qquad B=1\,,
\label{NAB}
\eeq
and the 4D scale factor obeys the following modified 
Friedmann equation: 
\beq
\left ( {{\dot R}\over R} \right)^2 \,=\, {8\pi\over 3}\,\gn\,
\rho\,-\, m_c\,{ {\dot R} \over R }\,.
\label{FriedmanNew}
\eeq
The $m_c\to 0$ limit of this equation is clearly incompatible
with Eq.~(\ref{Friedmann}) which is based on the
leading order approximation in the massive theory, but coincides with 
the result of massless gravity. 
This certainly implies that the Hubble parameter $\dot R/R$ is continuous 
in this limit --- the assertion we verify below by presenting the
exact solution of Eq.~(\ref{FriedmanNew}). 

We can absorb the parameters $m_c$ and $\gn$ in Eq.~(\ref{FriedmanNew})
by rescaling, 
\begin{eqnarray}
&& 
t=\frac{\tau}{m_c}\,,\qquad \qquad~~ \rho=\frac{3\,m_c^2}{32\pi \gn}\,\tilde
\rho
\,,\nonumber\\[2mm]
&&\left ( {{\tilde\rho^{\,\prime}}\over \tilde\rho} \right)^2=\, 
{9\over 4}\,\tilde
\rho+3\,{{\tilde\rho^{\,\prime}}\over \tilde\rho}\,,\qquad
\tilde\rho^{\,\prime}=\frac{{\rm d}\tilde\rho}{{\rm d}\tau }\,.
\end{eqnarray}
After introducing  the variable 
\beq
x\,\equiv1+\tilde\rho=1\,+\,{ 32\pi\,\gn\over 3\,m_c^2}\,\rho\,,
\label{X}
\eeq
the exact solution can be written in terms of elementary functions for
$\tau (x)$,
\beq
\frac{3}{2}\,m_c\,t\,=\,{  1\over \sqrt{x}-1 }\,
+\,{1\over 2} \,{\log} {\sqrt{x}+1\over \sqrt{x}-1}
 \,.
\label{xsolution}
\eeq

When $\tau =m_c\,t\gg 1$ we get for the scale factor,
\begin{equation}
R^3=\frac{8\pi \gn u}{m_c}\,t\left[1-\frac{\log(3m_c
t)+1}{3\, m_ct}+\ldots\right]\,.
\end{equation}
This unusual  (compare with the 4D Newtonian
cosmology in Eq.~(\ref{R3})) behavior is typical of a pure 
brane cosmology regime \cite{bdl} where one has 
$H^2 \propto \rho^2$ -- indeed, $\gn/m_c=1/(32\pi M_*^3)$ plays the
role of $\gn$ in the 5D world. It is only relevant to the late time cosmology, 
$t\gg 1/m_c$ --- the epoch where the Hubble parameter is small, 
$H\sim 1/t \ll m_c$, and the expansion enters the 5D regime, as analyzed in
\cite{Cedric}. Therefore, the 4D Newtonian cosmology is not applicable 
at this epoch.

For $\tau =m_c\,t\ll 1$ 
\begin{equation}
R^3=6\pi \gn u\, t^2\left[1-\frac{3}{4}\,m_ct+\ldots\right]\,.
\label{R3e}
\end{equation}
In correspondence with the discussed above difference of
Eqs.~(\ref{FriedmanNew}) and  (\ref{Friedmann}), 
 we see that $R^3$ at $m_c=0$ is different from the expression 
in Eq.~(\ref {R3}) 
which was obtained using the lowest tree-level approximation
by the same factor $3/4$ --- it contains $\gn$ instead of $\tgn$.
Note, that the exact expression for  $R^3$ is linear in $\gn$ --- 
no higher orders
are present.

The exact solution  considered above gives an explicit demonstration of
the nonperturbative continuity in the limit $m_c\to 0$.  
This continuity is not
uniform ---  for the given value of $t$ the parameter $m_c$ should be much 
smaller than $1/t$. This is the most strong constraint on the graviton mass
coming from cosmology, $ m_c\le H_0$, where $H_0\sim 10^{-42}~{\rm GeV}$ is
the present day Hubble parameter.

\section{An Interpolating Solution}

In this section we discuss a  cosmological solution 
found in \cite {ddgl} and show that it 
interpolates between the regimes with 4D and the 5D  tensor 
structures. 

Let us start with the brane action (\ref {action})
and in addition introduce in the theory a negative cosmological
constant on the brane $\Lambda_b$  and the matter density 
$\rho \ge |\Lambda_b|$ (we put pressure equal to zero
for simplicity). 
The time evolution of such a 4D brane 
universe is interesting, it evolves asymptotically to 
a static Minkowski space on the brane without any fine tuning 
\cite {ddgl}. The asymptotic form of the metric is as follows: 
\begin{equation}
ds^2 \,=\, -(1+b~|y|)^2 ~dt^2\,+ \,dx^idx_i\, + \,dy^2~, 
\label{asym}
\end{equation}
where the constant $b$ is
\begin{equation}
b~\equiv~|\Lambda_b| /4\M^3~.
\label{c}
\end{equation}
In fact, this is a solution to 
the equation 
\begin{equation}
\Rbu_{AB}\,-\,{1\over 2}\, G_{AB}\,\Rbu \, =\ {1\over
  2\M^{3}}~T_{AB}(x)\,
\delta(y)~,
\end{equation}
where the energy-momentum tensor on the brane is 
\begin{equation}
T_{\mu\nu} \,=\ {\rm diag}\,(0,-\Lambda_b,-\Lambda_b,-\Lambda_b )~,
\qquad T_{5\mu}=T_{55}=0\,,
\label{source1}
\end{equation}
i.e. $\rho +\Lambda_b\to 0$ in this limit.
To warrant the 4D behavior,  
 the induced 4D Ricci scalar on the brane was added in \cite {ddgl}.

The important thing is that the early cosmology of this model 
is standard, with no discontinuity in the Newton constant.
Indeed, the Friedmann equation is given in  (\ref {FriedmanNew}) where
$\rho$ should be substituted by $\rho+\Lambda_b$. The Newton constant on the 
right hand side of this equation is the conventional 4D gravitational
constant which reflects {\it no discontinuity}. 
This is  true as long as the early cosmology is concerned.

Let us now look at the late cosmology, more precisely at 
the form of the metric (\ref{asym}) to which the solution 
asymptotes. The metric on the brane is Minkowskian 
and static everywhere with only dependence on $y$.
For small values of $y$, which satisfy $b|y|\ll 1$, 
this metric can be obtained 
as a perturbation on the flat Minkowski space.
Indeed, for small perturbations (\ref {pert}) 
in the harmonic gauge (\ref {gauge}) 
we find Eq.~(\ref {basic}) 
with the energy momentum tensor defined in  (\ref {source1}).
This equation has the 5D tensor structure on the right hand side. 
Let us now notice that the energy-momentum tensor
(\ref{source1}) satisfies the relation:
\begin{equation}
T_{ij} - \frac{1}{3}~T\eta_{ij} =0\,,\qquad i,j=1,2,3\,.
\end{equation}
Therefore, the equation for $h_{ij}$ is simplified. 
This is completely due to the 5D tensor structure; in fact
had we have a 4D tensor structure, this would not be so.
Furthermore, the solution of equation (\ref{basic})  
in the gauge (\ref{gauge}) can be written in the following form:
\beq
h_{00}\, =\, -h_{55}\,=\,   
-\frac{|\Lambda_b|}{2~\M^3}|y|~, \qquad
h_{ij}\,=\, 0~, ~ \qquad h_{\mu 5}~=~0~.
\eeq 
One can indeed verify that this solution coincides in  
the first order with the exact solution (\ref{asym}). For this 
we perform the following  gauge  transformation of the exact 
solution (two different signs correspond to the two sides of the 
brane): 
\begin{equation}
y~ = ~{\rm sign}(z)~ \frac{1}{b}~ \left [\left(1+2b|z|+2b^2z^2\right)^
{1/2}\!-1\right ]~.
\end{equation}
After this the metric takes the form 
\begin{equation}
ds^2 \,=\ -(1+2b|z|+2b^2z^2)~dt^2 \,+ \,dx^i dx_i \,+\,
\frac{(1+2b|z|)^2}{1+2b|z|+2b^2z^2}~dz^2~,
\end{equation}
which in the leading order coincides with the perturbative solution. 

Therefore, we conclude that 
the cosmological solution of Ref. \cite {ddgl}
does  indeed provide an explicit example 
with both asymptotic regimes:  at small distances
(small Hubble radius) the behavior is 4-dimensional with the 
4D tensor structure, whereas at large distances (large Hubble radius) 
the behavior has the 5D tensor structure. 
In this sense the solution discussed above  
captures  the important  features of  a Schwarzschild solution of 4D massive 
theory; this is not surprising  since it is 
asymptotically (in time) Minkowski on the brane.

\section{Discussions and Conclusions}

We discussed  a nonlinear five-dimensional generally covariant
model which resembles many crucial properties of 
a massive graviton in four-dimensions.
The  mass discontinuity is present
in the lowest tree-level approximation, 
however, this approximation breaks down for 
the vanishing graviton mass and all the tree-level graphs should be 
taken into account. The resulting  expression of the 
nonperturbative classical calculation is  continuous in the graviton 
mass.  Thus, there is no mass discontinuity in the 
full classical theory.  

There are three extra 
degrees of freedom in the massive (or 5-dimensional) 
theory compared to the massless one. 
Among these degrees of freedom  only the helicity 0 state (the graviscalar) 
has a nonzero coupling to 4D matter.  
However, this coupling tends to zero in full classical theory  
as the graviton mass (or $m_c$ in the 5D example) 
vanishes. Thus, all the extra degrees of freedom  
decouple in the massless limit.
  
The interesting issue which we did not discuss
in the paper is the emission of a helicity 0 gravitons.
Based on our  observations and using the unitarity arguments
we expect that the nonperturbative 
amplitudes of the radiation of the helicity 0 state 
by 4D  matter fields also vanish 
with the graviton mass,  while they are non-vanishing  
in the lowest tree-level approximation as was shown in
Ref. \cite {PP}.

In the small mass limit the extra degrees of freedom 
of a massive theory form an independent sector which 
decouples from our matter as the graviton mass goes to zero. 
These degrees of freedom do interact with each other; 
moreover, in perturbation theory these interactions are 
singular in the limit $m_g\to 0$.
Certainly, on top of the classical 
effects there is an issue of quantum loops
which we did not discuss in the present work.
However, the loop effects are suppressed and 
most likely  they cannot be disentangled in 
existing measurements.

\vspace{0.5cm}   

{\bf Acknowledgments}
\vspace{0.1cm} \\

We would like to thank  L.~Blanchet, P.~A.~Grassi, 
I.~Kogan, M.~Porrati and M.~Shaposhnikov for 
useful discussions. 
The work of C.D. is supported by a David and Lucile  
Packard Foundation Fellowship for  Science and Engineering
and by NSF grant PHY-9803174. The work of G.D. is supported in part 
by a David and Lucile  
Packard Foundation Fellowship for  Science and Engineering,
by Alfred P. Sloan foundation fellowship and by NSF grant 
PHY-0070787. G.G. and A.V.
are supported by  DOE grant DE-FG02-94ER408.

\end{document}